\documentclass[%
  aps,
  pra,               
  longbibliography,  
  amsmath,
  amssymb,
  reprint,           
  twocolumn,         
  floatfix           
]{revtex4-2}

\usepackage{amsmath,amssymb}
\usepackage{graphicx}
\usepackage{xspace,soul,color}

\usepackage{xcolor}

\definecolor{myred}{rgb}{1,0.8,0.8}

\definecolor{mycyan}{rgb}{0.5,0.92,1.0}

\definecolor{mycyan}{rgb}{0.5,0.92,1.0}
\definecolor{mygreen}{rgb}{0.56,0.93,0.56}

\definecolor{myhl}{rgb}{1.0,0.98,0.56}
\sethlcolor{myhl}

\usepackage{graphicx} 
\usepackage{dcolumn}  
\usepackage{bm}       
\usepackage{subfigure}

\usepackage[T1]{fontenc}
\usepackage{lmodern}

\usepackage{hyperref}

\makeatletter

\renewcommand{\@fnsymbol}[1]{%
  \ifcase#1\relax
  \or \dag
  \or \ddag
  \fi
}
\makeatother

\begin{document}

\preprint{APS/123-QED}

\title{Spectral Distribution of one-dimensional Photonic Quasicrystals:\\ The Role of Irrational Numbers}

\author{Hui Quan}

\author{Wei Si}  
\thanks{Contact author: siwei@xtu.edu.cn}  

\author{Kai Jiang} 
\thanks{Contact author: kaijiang@xtu.edu.cn}  

\affiliation{
Hunan Key Laboratory for Computation and Simulation in Science and Engineering, Key Laboratory of Intelligent Computing and Information Processing of Ministry of Education, School of Mathematics and Computational Science, Xiangtan University, Xiangtan, Hunan 411105, China
}



\begin{abstract}

In this paper, we construct a one-dimensional photonic quasicrystal by combining two incommensurate spatial harmonics, where the ratio of their periods is the irrational number $\beta$.
We evaluate the photonic quasicrystal accurately by a generalized spectral method that embeds the quasiperiodic structure into a higher-dimensional periodic system.
We study the spectral distribution of one-dimensional photonic quasicrystals and find some interesting phenomena.
As the computational resolution $N$ increases, there are more eigenvalues within finite frequency bandwidths, and the maximum localization always occurs at spectral gap edges for states near index $N+1$.
By varying $\beta$ within the range of $(0, 1)$, we present a butterfly-shaped spectral structure with abundant band gaps.
We find that the spectral structure factor $Q$ (defined as $I_{mg}/N$, where $I_{mg}$ is the maximum gap index) exhibits different linear patterns as $\beta$ changes: $Q = 1 - \beta$ when $\beta < \beta_c$, while $Q = \beta$ when $\beta > \beta_c$, where $\beta_c \approx 0.424$ is the transition point.
This linear relationship holds robustly in the strong quasiperiodic regime ($\beta$ away from 0 or 1) and is independent of the specific type of irrational number used.
The relationship disappears (weak quasiperiodic regime) near $\beta=0$ or $\beta=1$.
It demonstrates that the intrinsic spectral properties of one-dimensional photonic quasicrystals are fundamentally governed by the magnitude of the irrational parameter $\beta$.


\end{abstract}

\maketitle

\section{Introduction}
\label{sec:intro}

Photonic quasicrystals (PQCs) are optical structures that exhibit long-range order but lack translational symmetry, closely related to irrational numbers.
Compared with traditional photonic crystals, PQCs lack translational symmetry, which provides rich intrinsic physical properties and vast prospects for electromagnetic manipulation; thus, PQCs have received widespread attention.
In recent years, extensive research has been carried out on PQCs\,\cite{vardeny2013optics, jeon2017intrinsic, che2021polarization, zhang2024non, yang2024observation, lin2025photonic}, revealing numerous surprising phenomena, such as disorder-enhanced transport\,\cite{levi2011disorder}, localization of light\,\cite{sinelnik2020experimental, wang2024observation, ivanov2025observation}, nonlinear optical frequency conversion\,\cite{lifshitz2005photonic, tang2019quasicrystal}, multifractality of states\,\cite{jagannathan2021fibonacci, reisner2023experimental}, and so on.
Among these, the localization phenomenon refers to the situation where light propagation in a quasiperiodic medium is affected by complex interference and scattering, resulting in the confinement of light waves to local spatial regions\,\cite{levi2011disorder}. 
The localization of light has significant practical value and can be applied in various fields, such as for terahertz lasers\,\cite{vitiello2014photonic,jia2025laser}, topologically protected light transmission\,\cite{bandres2016topological,kasture2014plasmonic}, and the formation of high quality factor nanocavities\,\cite{vasco2019exploiting}.

There has been extensive research on the spectral structures of various quasiperiodic systems\,\cite{liu2021localization, hege2022finding, zhu2024localization, comi2024some}.
The spectral distribution of these quasiperiodic systems exhibits a variant of the well-known "Hofstadter butterfly" structure\,\cite{hofstadter1976energy}, which possesses fractal and self-similar characteristics and contains abundant spectral gaps\,\cite{pal2019topological, xia2020topological, marti2021edge, wang2024non, davies2023graded}.
Since PQCs are related to irrational numbers, a natural question arises: What role do irrational numbers play in the butterfly graph of PQCs?
Considering the complexity of coupling may mask the underlying laws in high-dimensional systems, we focus on the 1D PQC with a single irrational number, where the variation of the irrational number directly controls the energy gap distribution, forming the fractal characteristics of the butterfly graph.

It requires a numerical method to evaluate PQCs accurately and efficiently.
A common method uses a periodic structure with a supercell to approximate the target quasicrystal structure, so-called periodic approximation method\,\cite{jiang2025approximation}.
This method inevitably leads to approximation errors, and these errors cannot be eliminated unless the computational box is the full space\,\cite{jiang2015stability, jiang2024numerical}.
Another method, the projection method (PM)\,\cite{jiang2014numerical}, fundamentally avoids the impact of approximation errors.
Based on the fact that a quasicrystal can be embedded into a high-dimensional crystal structure, the PM computes the crystal structure accurately and efficiently in a high-dimensional unit cell and then projects it back to the original physical space.
Until now, the PM has been successfully applied to the computation of various quasiperiodic systems\,\cite{wang2022effective, cao2021computing, jiang2022tilt, jiang2025irrational, jiang2025projection, si2025designing}.

In this paper, we use the PM to obtain the exact spectral distribution of 1D PQCs and further explore the role of irrational numbers within it.
Under a quasiperiodic optical lattice (QOL) composed of two spatial harmonics (the ratio of their periods is the irrational number $\beta$), we calculate and analyze the eigenvalues and eigenstates of 1D PQCs with different $\beta$ and different resolution $N$.
We summarize the filling rule of eigenvalues and the index rule of maximum localization eigenstate as the resolution $N$ increases.
Fixing the resolution, we plot a butterfly-shaped spectral structure by varying the irrational number $\beta$ in (0,1).
What we are concerned about is the maximum band gap, and we find that its index has a linear relationship with $\beta$ in a strong-quasiperiodic region ($\beta$ is far from 0 and 1).
This linear relationship only depends on the magnitude of $\beta$, and is independent of the type of $\beta$ and the size of $N$.
This discovery helps us better understand the significant role of irrational numbers in the control of PQCs.

%

\section{Model and method}
\label{sec:model}

The system under consideration exhibits a varying lattice structure only in the $x$-direction, while being uniform in other two directions. For such a 1D system, the propagation of electromagnetic waves can be described by the following scalar eigen equation \,\cite{joannopoulos2008molding, vaidya2023reentrant}, which is derived from Maxwell's equations, 
\begin{equation}
	-\frac{d }{d x}\left(\frac{1}{\varepsilon(x)}\frac{d }{d x}H(x)\right)=\left(\frac{\omega}{c}\right)^2\!H(x),
	\label{eq:mode_l}
\end{equation}
where $\omega$ represents the angular frequency of the electromagnetic wave, $c$ denotes the speed of light in vacuum, and $\varepsilon(x)$ represents the dielectric function that varies along the $x$-direction. $H(x)$ is the scalar magnetic field propagating along the $x$-direction, with its field distribution strongly influenced by the lattice structure of the dielectric function $\varepsilon(x)$.

We consider a QOL distribution to set up the dielectric function, which can be used to generated a 1D PQC.
We express the QOL distribution as the superposition of two incommensurate spatial harmonics,
\begin{equation}
	\begin{aligned}
        \varepsilon^{-1}(x)=0.5\left[\cos(x)+\cos(\beta x) \right]+1,
    \end{aligned}
    \label{eq:epsilon_1} 
\end{equation}
where $\beta$ is an irrational number representing the ratio of the periods of the two harmonics. Without loss of generality, we assume $0<\beta<1$. Similar optical lattice setups can be found in many studies of quasiperiodic systems\,\cite{zhu2024localization, gao2023pythagoras, yu2024observing,kartashov2025localized}.


An accurate and efficient approach to studying PQCs is the PM\,\cite{jiang2014numerical}.
The PM embeds the PQC into a high-dimensional periodic system (the corresponding space is denoted by $Z$), which can be efficiently calculated through the Fast Fourier Transform, and then recovers the PQC by projecting it back into the original physical space (denoted by $X$).
For a 1D PQC generated by the QOL\,\eqref{eq:epsilon_1}, the projection matrix is $\bm{P} = [1, \beta]$, which determines the projection direction in the 2D superspace.
The irrational number $\beta$ ensures the fact that the straight line corresponding to the projection direction is dense after being mapped back to the superspace unit cell\,\cite{janot2012quasicrystals, jiang2024accurately}.
We carried out the following generalized Fourier expansion of the dielectric function,
\begin{equation}
	\varepsilon^{-1}(x) := \mu(x) = \sum_{\mathbf{k}\in \mathcal{L}} \hat{\mu}(\mathbf{k}) e^{i(\bm{P}\mathbf{k})\cdot x},
	\label{eq:expansion_1}
\end{equation}
where the reciprocal lattice $\mathcal{L} = \{ \mathbf{k} = \sum_{i=1}^2 k_i \mathbf{b}_i \ |\ k_i \in \mathbb{Z} \}$ is composed of integer linear combinations of the reciprocal basis vectors $(\mathbf{b}_i)_{2\times 1}$ in 2D superspace, and $\hat{\mu}(x)$ is the Fourier coefficient.
In this superspace, it is natural to apply Bloch's theorem to the eigenfunction\,\cite{rodriguez2008computation,davies2025convergence}, i.e., $H(z)=e^{i\bm{k_z}\cdot z}h(z)$, where $\bm{k_z}$ and $h(z)$ are the 2D wave vector and periodic function, respectively.
Thus the generalized Fourier expansion of the eigenfunction is
\begin{equation}
	H(x) = e^{ik_x\cdot x}\sum_{\mathbf{k}\in\mathcal{L}}\hat{H}(\mathbf{k})e^{i(\bm{P}\mathbf{k})\cdot x},
	\label{eq:expansion_2}
\end{equation}
where $k_x$ is the projection of the wave vector $\bm{k_z}$ in physical space $X$.
By substituting Eq.\,\eqref{eq:expansion_1}-\eqref{eq:expansion_2} into Eq.\,\eqref{eq:mode_l} and simplifying, we obtain the following equation, 
\begin{equation}
	\begin{aligned}
    \sum_{\mathbf{k^{\prime}} \in\mathcal{L} }\left[(k_x + \bm{P}\mathbf{k})\cdot(k_x + \bm{P}\mathbf{k^{\prime}})\right]\hat{\mu}&(\mathbf{k}-\mathbf{k^{\prime}})\hat{H}(\mathbf{k^{\prime}})\\
    &=\left(\frac{\omega}{c}\right)^2\!\hat{H}(\mathbf{k}).
    \end{aligned}
    \label{eq:EQU5}
\end{equation}

This corresponds to an eigenvalue problem in the 2D reciprocal space, which is linked to the 1D physical space $X$ through the projection matrix $\bm{P}$.
For a fixed irrational number $\beta$, we explicitly assemble Eq.\,\eqref{eq:EQU5} into matrix form and numerically solve to obtain a sequence of eigenvalues $\{(\omega_n/c)^2\}$, which we refer to as the eigenvalue spectra.

Traversing the projected wave vector $k_x$ actually corresponds to solving the band structure of 1D PQCs, which we do not intend to do in this study.
Since there is no strictly defined Brillouin zone in quasicrystals, here we only consider the wave vector at the $\Gamma$ point, i.e., $k_x = 0$.

In numerical calculations, we truncate the Fourier expansions Eq.\,\eqref{eq:expansion_1}-\eqref{eq:expansion_2} to a finite number of terms, i.e.,
$\mathcal{L}_N = \{ \mathbf{k} = \sum_{i=1}^2 k_i \mathbf{b}_i \ |\ k_i \in \mathbb{Z},\ -N/2 \leq k_i < N/2 \}$, where $N$ represents the number of Fourier modes as well as the computational resolution in the 2D unit cell.
The straight line in the projection direction densely fills the 2D unit cell, indicating that, as we increase the resolution $N$ in numerical calculations, more details within the PQCs will be revealed.

\section{Numerical results and discussions}
\label{sec:rslt}     

\subsection{Eigenvalues and eigenstates in 1D PQCs}

\begin{figure}[!htbp]
    \centering
    \includegraphics[scale=0.56]{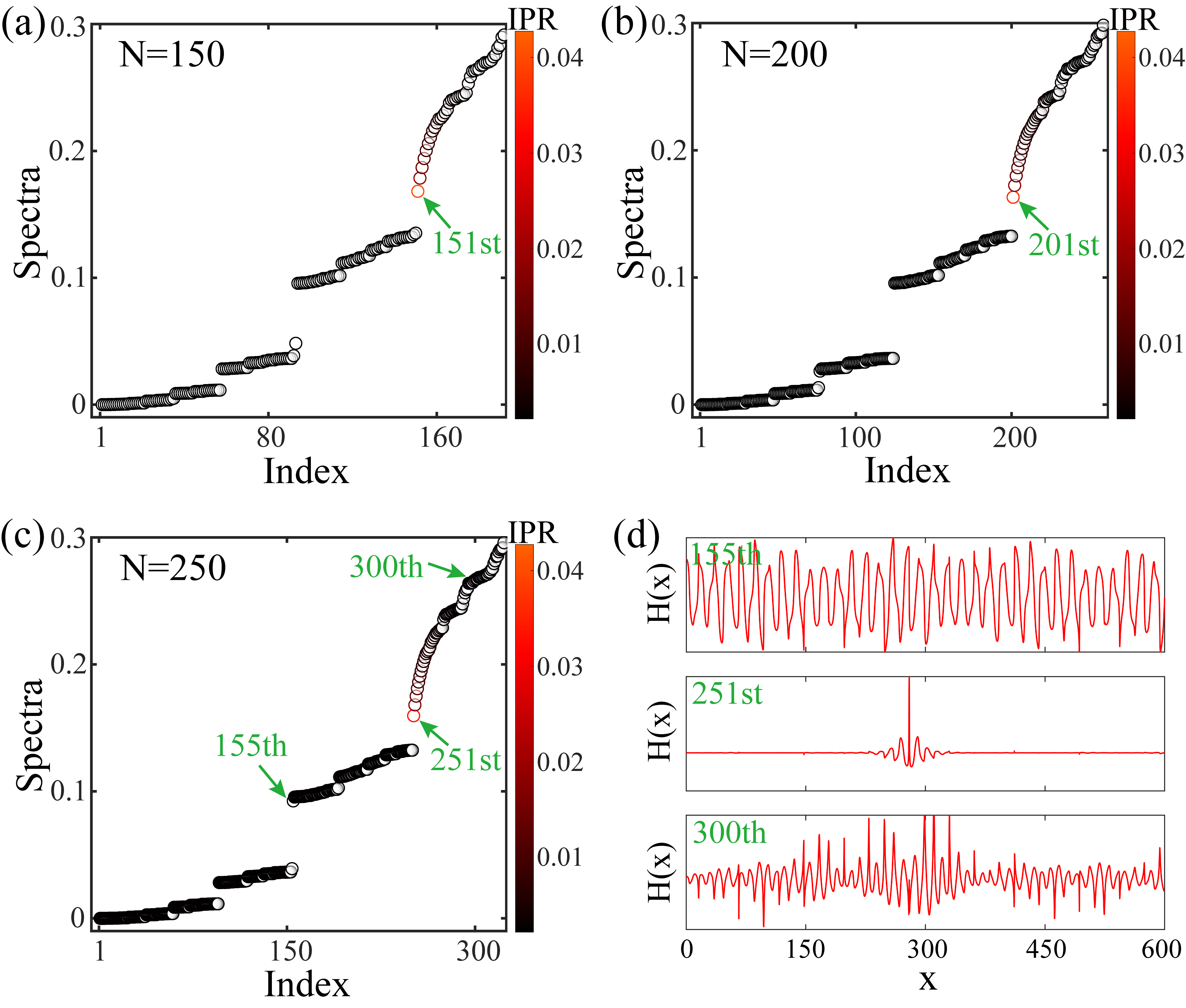} 
    \caption{
    (a)-(c) The eigenvalue spectral structure corresponding to the golden ratio $(\sqrt{5} - 1)/2$ at calculation resolutions of $N=150$, $200$ and $250$. As the resolution N increases, the number of spectra within a given frequency bandwidth also increases. (d) The field $H(x)$ distributions of the three eigenstates at a resolution of $N = 250$ in physical space $X$, which exhibit extended, localized, and extended characteristics, respectively.
    }
    \label{fig:Fig1} 
\end{figure}

The eigenvalue spectra of the PQCs (eigenvalue sequence $\{(\omega_n/c)^2\}$ solved by Eq.\,\eqref{eq:mode_l} ), are quite different from those of periodic photonic crystals.
For photonic crystals, only a finite number of eigenvalues exist within a finite frequency bandwidth \cite{bao2001mathematical}, which is determined by the fundamental properties of the elliptic operator in Eq.\,\eqref{eq:mode_l} under periodic boundary conditions\,\cite{gohberg2012basic}.
In contrast, PQCs allow an infinite number of eigenvalues\,\cite{rodriguez2008computation,davies2025convergence}.
As shown in Fig.\,\ref{fig:Fig1}(a)-(c), the number of eigenvalue spectra within a finite frequency bandwidth also increases as the computational resolution $N$ increases.
This is because the elliptic operator, when elevated to the 2D superspace, lacks derivative constraints in the perpendicular space $Y$, wich satisfies the relationship $Z = X \oplus Y$, causing the operator to no longer be compact, which leads to a continuous spectrum instead of discrete eigenvalues\,\cite{rodriguez2008computation,gao2025reduced}.
More intuitively, the absence of derivative constraints makes any phase shift $e^{ik_yy}$ in the perpendicular space $Y$ direction permissible.



To further explore the eigenstate properties, we compute the IPR as a measure of localization, with values represented by color in Fig.\,\ref{fig:Fig1}(a)-(c). 
The formula for calculating IPR is
\begin{equation}
	\begin{aligned}
	\text{IPR} = \frac{\sum_{j} \vert \mathcal{H}_{n,j} \vert^4}{\left( \sum_{j}  \vert \mathcal{H}_{n,j} \vert^2 \right)^{2}},
    \end{aligned}
    \label{eq:IPR}
\end{equation}
where $\mathcal{H}_{n,j}$ represents the field distribution of the $n$-th state within the 2D superspace unit cell, with $j$ indicating the computational grid point. 
This means that we calculate the IPR directly from the field distribution in the 2D unit cell, rather than projecting it into physical space $X$ and truncating it to a finite length for computation. 
Because the 2D superspace unit cell captures all the information of the field distribution in full space, while the latter only contains a local segment of the infinite field distribution.
As the eigenstate index increases, the IPR first stays low (black), rises sharply, then gradually decreases (from red to black), and finally drops back to a low level (black), representing a transition from extended to localized to extended states. 
Moreover, the state with the maximum IPR always appears at the edge of the eigenvalue spectra gap, corresponding to the $(N+1)$-th state. 
Note that the eigenstate transition here does not involve parameter tuning, which is different from the Reentrant Delocalization Transition in recent studies\,\cite{vaidya2023reentrant, roy2021reentrant, wang2025coexistence}.

Fig.\,\ref{fig:Fig1}(d) shows the eigenstates corresponding to three different indices in the physical space $X$, with a resolution of $N=250$. 
The field distributions of the $155$th and $300$th states oscillate continuously with the physical space coordinate $x$, exhibiting wave propagation characteristics. 
While the $251$st state shows a sharp oscillation peak in a local region, indicating suppressed wave propagation. 
This localized state seems to appear only at the edge of spectral gaps for specific frequencies, with eigenstates quickly returning to extended states beyond this frequency.
This could be related to the resonance of electromagnetic waves. 
Specifically, the QOLs related to irrational numbers disrupt the free diffusion of waves, making propagation more complex. 
When electromagnetic waves of a specific frequency resonate with the QOLs, energy is concentrated in local regions, leading to the emergence of localized states.

Next, we want to explore the impact of varying the sizes of irrational numbers on the eigenvalues and eigenstates of 1D PQCs. 
For this purpose, we introduce the rational linear transformation to generate an irrational sequence, which allows us to freely adjust the size of irrational numbers while specifying their type. 
Specifically, we select an irrational number $\beta_0$ and apply the linear transformation $\beta = a\beta_0+b$ to it, where $a$ and $b$ are rational numbers, resulting in a new irrational number. 
By varying the values of $a$ and $b$, we can generate a sequence of irrational numbers $\{\beta_n\}$, which is of the same type as $\beta_0$. 
For example, the golden ratio $(\sqrt{5} - 1)/2$ can be generated through a rational linear transformation of $\sqrt{5}$ and is essentially the same type of irrational number. 
Without loss of generality, we restrict the irrational number sequence $\{\beta_n\}$ to within $(0, 1)$.

\begin{figure}[!htbp]
    \centering
    \includegraphics[scale=0.5]{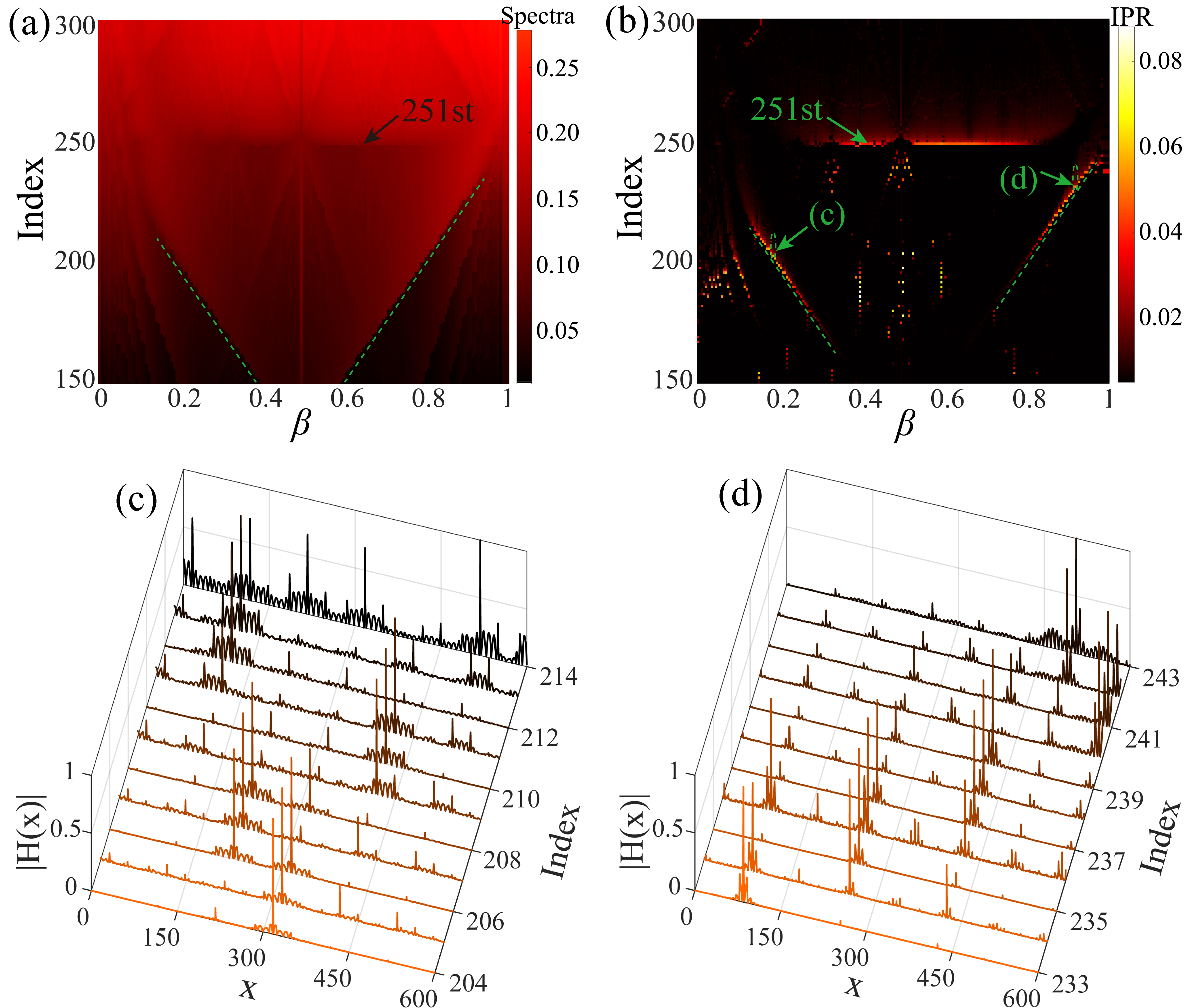} 
    \caption{
    (a) Index of eigenvalue spectra versus irrational number $\beta$, where color represents the value of eigenvalue. (b) The IPR distribution corresponding to (a). The maximum spectral gaps correspond to the bright IPR stripes, marked by arrows and green dashed lines. (c)(d) The field amplitude $|H(x)|$ distribution of the 11 eigenstates marked by green arrows in (b), which gradually revert to the extended states as the index increases.
    }
    \label{fig:Fig2} 
\end{figure}

By applying the rational linear transformation, we obtain an irrational sequence of the same type as the golden ratio $(\sqrt{5} - 1)/2$, which is generated by multiplying it by a rational number $a$ and adding another rational number $b$. 
We traverse this sequence and compute the corresponding eigenvalue spectrum at a resolution of $N = 250$, sorting the spectrum in index order.
We select a segment of the spectra of interest and represent its values with color, resulting in Fig.\,\ref{fig:Fig2}(a). 
In this figure, regions with noticeable color contrast indicate the existence of spectral gaps. 
The large spectral gaps on either side are marked by green dashed lines, and the gap in the middle region by a black arrow.
As shown in Fig.\,\ref{fig:Fig2}(b), we compute the corresponding IPR phase diagram using Eq.\,\eqref{eq:IPR} and visualize its values with color. 
The bright IPR stripes are marked by green dashed lines and arrows. 
The bright stripes highly coincide with the spectral gap positions, except for the puzzling vertical bright IPR stripes between the two green dashed lines. 
This suggests that localized states tend to appear at the edges of the spectral gaps, consistent with previous analysis.
Moreover, the brightest IPR stripes in the middle region correspond to the $(N+1)$-th state.
It indicates that for irrational numbers in the middle of the interval $(0, 1)$, the index of the most localized eigenstate may be related to the computational resolution $N$.

We then select $11$ eigenstates on each side of the bright IPR stripes, marked with green arrows.
We plot the distribution of the field amplitude $|H(x)|$ in physical space, using warm and cool colors to distinguish between localized and extended states, as shown in Fig.\,\ref{fig:Fig2}\,(c)-(d). As the index increases, the eigenstates gradually revert to extended states. 


\subsection{The influence of irrational numbers on spectral distributions}

\begin{figure}[!htbp]
    \centering
    \includegraphics[scale=0.50]{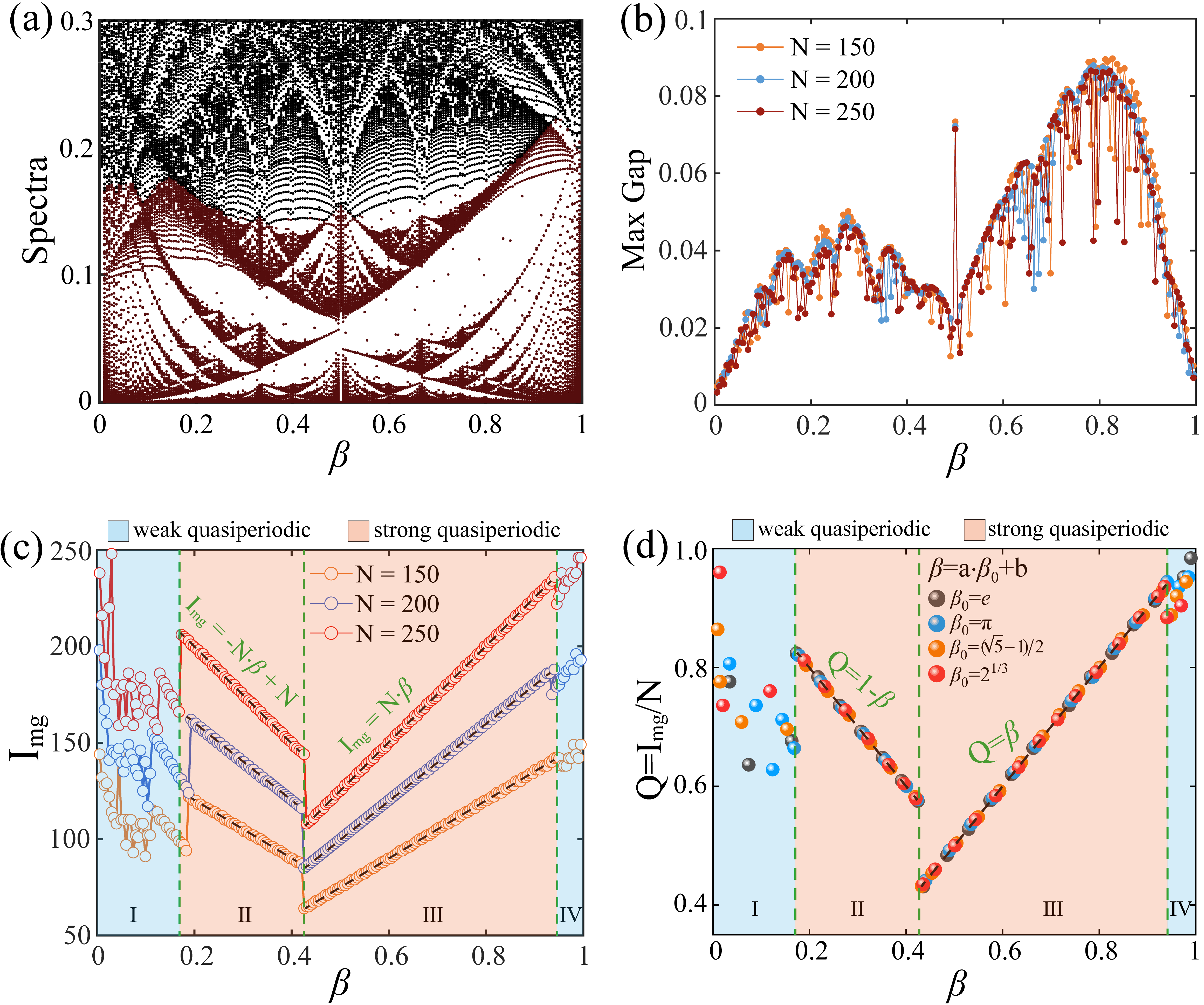}
    \caption{
    (a) The butterfly-shaped spectral structure with a calculation resolution of $N = 250$, where the first $N$ eigenvalue spectra are marked in dark red. (b) The maximum gap distribution is extracted from the first $N$ eigenvalue spectra at different computational resolutions. 
    (c) The index of the maximum gap ($I_{mg}$) extracted at different resolutions $N$.
    The interval (0,1) is divided into regions I, II, III, and IV, which are marked with light orange (strong quasiperiodic) and light blue (weak quasiperiodic).
    (d) The factor $Q$ ($Q = I_{mg}/N$) maintains a linear relationship with $\beta$ in the strong quasiperiodic region, while it disappears in the weak quasiperiodic region. This phenomenon is independent of the type of irrational number.
    }
    \label{fig:Fig3}
\end{figure}

Much interest has arisen in the distribution of spectral gaps in PQCs.
In this subsection, our main objective is to investigate the role that irrational numbers play in the spectral distribution of PQCs.

We start with the irrational number sequence (generated by the golden ratio), sort the eigenvalue spectra by magnitude, and mark the first $250$ eigenvalues in dark red, as shown in Fig.\,\ref{fig:Fig3}(a).
It exhibits a typical butterfly-shaped spectral structure with abundant spectral gaps.
As the parameter $\beta$ is tuned, the gaps in the spectra shift up and down, opening or closing accordingly.
This butterfly-shaped spectral structure has been reported in many quasiperiodic systems\, \cite{pal2019topological,xia2020topological,marti2021edge,wang2024non,davies2023graded}, and typically exhibits self-similarity and fractal structures.
If we use a higher resolution $N$ for computation, more eigenvalues will be revealed, and the intrinsic properties of the spectral structure remain unchanged.
Here, we focus on the maximum gap in the first $N$ eigenvalue spectra.
Fig.\,\ref{fig:Fig3}(b) shows the distribution of the maximum gap as a function of $\beta$ for resolutions $N = 150$, $200$, and $250$.
The random fluctuations in the data points are caused by the unexpected eigenvalues that occur in the spectral gaps under the limited resolution. These eigenvalues obscure the original maximum gap, and the corresponding eigenstates are referred to as "spurious modes" in previous studies\, \cite{rodriguez2008computation,davies2025convergence}.

Now, let us focus on the index of the maximum gap, which is denoted as $I_{mg}$.
Fig.\,\ref{fig:Fig3}(c) illustrates the distribution of $I_{mg}$ as a function of $\beta$ at different resolutions.
Based on its distribution characteristics, we divide the (0,1) interval into four distinct regions, which are labeled as I, II, III, and IV, respectively.
In regions II and III, there is a linear relationship between $I_{mg}$ and $\beta$. We refer to the two regions as the strong quasiperiodic regions, which are displayed in light orange.
Through data fitting, we are surprised to find that at different resolutions, $I_{mg}$
all follows the linear equation $I_{mg} = -N\beta + N$ in region II but satisfy $I_{mg} = N\beta$ in region III.
More test results at different computational resolutions $N$ all exhibit such a linear relationship.
In regions I and IV, the linear relationship no longer holds. We refer to the regions as weak quasiperiodic regions, which are represented by light blue.

To further explore whether this linear relationship is just a coincidence, we divide $I_{mg}$ by $N$ at different resolutions and define it as $Q$, i.e., $Q = I_{mg} / N$. 
This is actually an invariant, that is, the data in strong quasiperiodic regions II and III at any resolution can be represented by the linear relationships $Q = 1 - \beta$ and $Q = \beta$, respectively.
Furthermore, we also find that this linear relationship is not affected by the type of irrational numbers. 
We test various types of irrational numbers, such as the natural constant $e$, the mathematical constant $\pi$, the golden ratio $(\sqrt{5}-1)/2$, $2^{1/3}$, $\sqrt{2}$, $\sqrt{3}$, and so on, all of which follow this relationship in regions II and III. 
In Fig.\,3(d), we present the first four of these. While in weak quasiperiodic regions I and IV, this linear relationship no longer holds. 
Overall, the linear relationship within the strong quasiperiodic region reveals an interesting fact, namely that the spectral distribution of the 1D PQCs is related to the irrational number $\beta$. Specifically, the defined spectral structure factor $Q$ shows a linear relationship with $\beta$, which depends on the magnitude of the irrational number but is independent of its type. 

\begin{figure}[!htbp]
    \centering
    \includegraphics[scale=0.58]{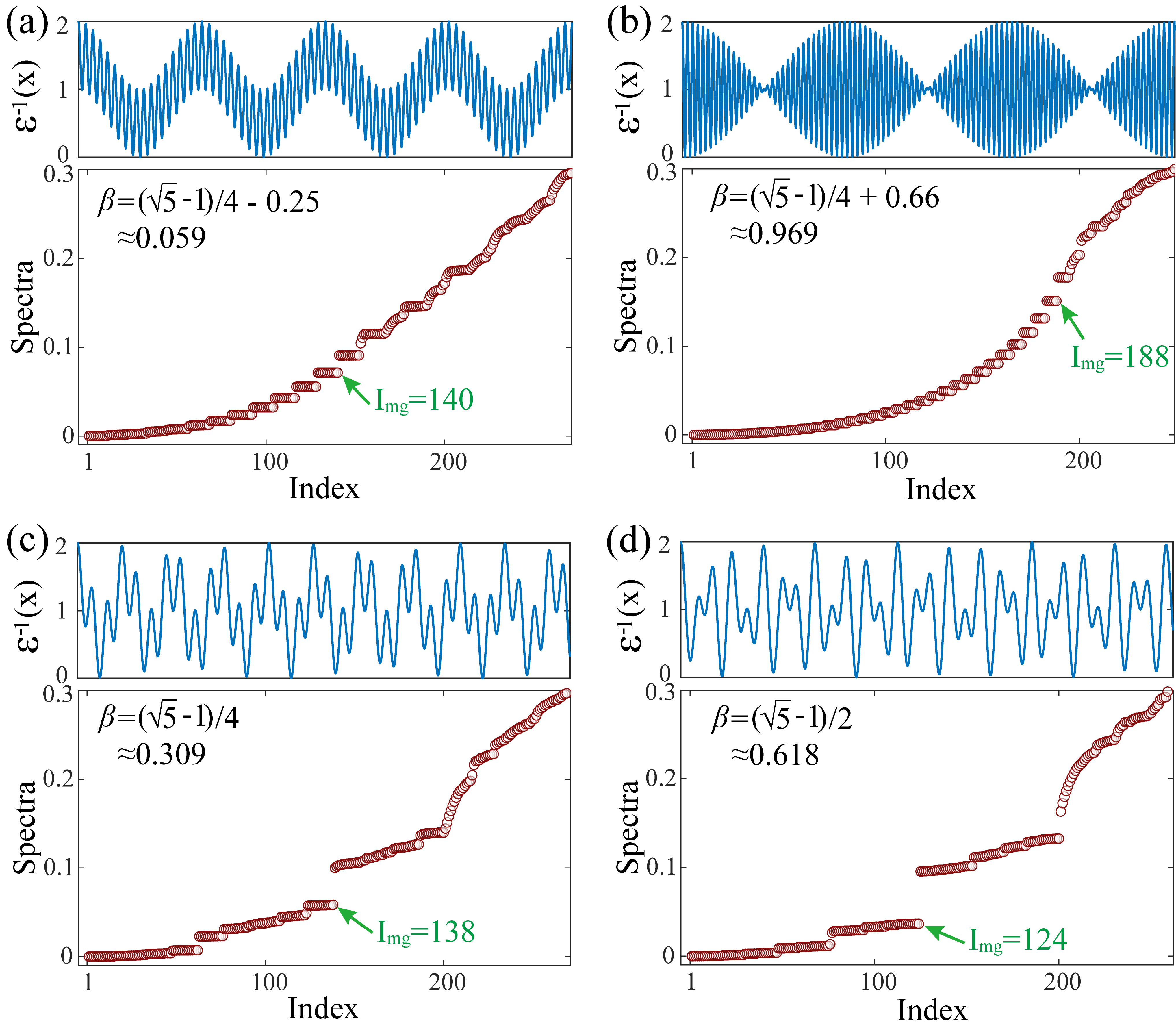}
    \caption{
    The lattice potentials (blue curves) and eigenvalue spectra (dark red scattered points) corresponding to the irrational numbers in regions I–IV. (a)(b) For the irrational numbers in regions I and IV (close to 0 and 1, respectively), the envelope of the lattice potential $\varepsilon^{-1}(x)$ exhibits a periodic-like feature, and the eigenvalue spectra contain many short and small gaps. (c)(d) For irrational numbers in regions II and III (far from 0 and 1), the lattice potential $\varepsilon^{-1}(x)$ is the strong quasiperiodic, and the eigenvalue spectra show obvious gaps. The $I_{mg}$ is marked by the green arrow. 
    }
    \label{fig:Fig4}
\end{figure}

\begin{figure*}[htbp]
    \centering
    \includegraphics[scale=0.72]{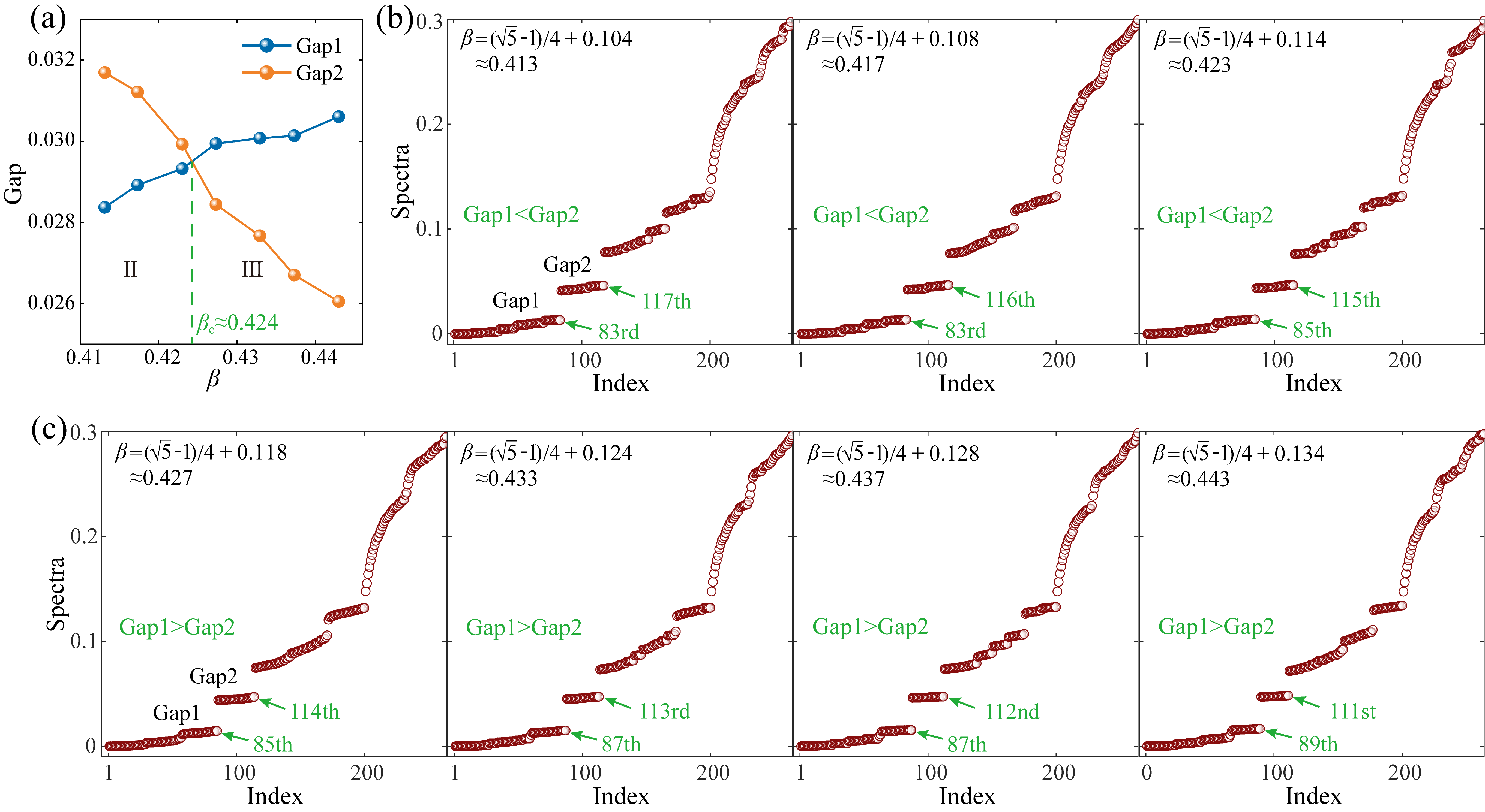}
    \caption{
    (a) Evolution of the first two maximum eigenvalue spectra gaps (Gap1, Gap2) for irrational numbers near the critical point $\beta_c$ of regions II and III. (b) The irrational number to the left of the critical point ($\beta<\beta_c$) satisfies Gap1$<$Gap2. (c) The irrational number to the right of the critical point ($\beta>\beta_c$) satisfies Gap1$>$Gap2. The indices of Gap1 and Gap2 are marked by green arrows.
    }
    \label{fig:Fig5}
\end{figure*}

A natural question arises: why is there no such linear relationship in regions I and IV?
By applying a rational linear transformation to the golden ratio, we obtain four distinct irrational numbers (corresponding to regions I–IV respectively) and observe their corresponding lattice potentials (blue curves) and eigenvalue spectra (dark red scattered points), as shown in Fig.\,\ref{fig:Fig4}.
The expression of the lattice potential is given by Eq.\,\eqref{eq:epsilon_1}, which consists of two incommensurate periodic lattices, $\cos(z)$ and $\cos(\beta z)$, with fundamental periods of $2\pi$ and $2\pi/\beta$, respectively.
When $\beta$ approaches $0$ or $1$, one lattice dominates, or both are roughly equivalent, in which case the irrational number has a weak control over the system, corresponding to the weak quasiperiodic.
Ignoring local details, the profile of the lattice potential exhibits features similar to those of a periodic lattice, and the corresponding eigenvalue spectra structure shows many short and small gaps, as shown in Fig.\,\ref{fig:Fig4}(a)(b).
When the irrational number $\beta$ is far from $0$ and $1$, it exerts a strong control over the lattice potential, leading to the strong quasiperiodic, with the eigenvalue spectra showing obvious gaps, as shown in Fig.\,\ref{fig:Fig4}(c)-(d).
It is clear that the value of the irrational number strongly influences the distribution of the lattice potential and the structure of the eigenvalue spectra.
We suggest that it is the strength of its control over the quasiperiodic system that gives rise to such distinct outcomes.

Finally, we discuss the behavior near the critical point in regions II and III.
By applying rational linear transformations to the golden ratio, we obtain irrational numbers with varying magnitudes near the critical point and observe the evolution of the maximum and submaximum gaps in the corresponding eigenvalue spectra.
We denote the first two maximum gaps as Gap1 and Gap2, with Gap1 corresponding to the smaller index and Gap2 to the larger index.
In Fig.\,\ref{fig:Fig5}(a), we present the trends of Gap1 and Gap2 as a function of the irrational number $\beta$.
It can be observed that Gap1 exhibits an increasing trend with $\beta$, while Gap2 shows a decreasing trend, and the critical value $\beta_c = 0.424$ is identified at their intersection. 
For irrational numbers on the left of the critical value $(\beta < \beta_c)$, the maximum gap corresponds to Gap2; for those on the right of it $(\beta > \beta_c)$, it corresponds to Gap1.
Their eigenvalue spectra are shown in Fig.\,\ref{fig:Fig5}(b)-(c), respectively.
As $\beta$ changes, the positions of the maximum and submaximum gaps swap between Gap1 and Gap2, resulting in a discontinuity in the value of $I_{mg}$.

\section{Conclusion}

Based on an accurate numerical method to compute PQCs, we study the eigenvalues and eigenstates of 1D PQCs and find some interesting phenomena.
There are an infinite number of eigenvalues within a given frequency bandwidth, which is distinguished by periodic photonic crystals.
We use the 2D superspace field distribution to compute the IPR of eigenstates, and find that the localized states tend to appear at the edges of the spectral gaps. For irrational numbers in the middle part of the interval $(0, 1)$ (not near $0$ or $1$), the most localized state always corresponds to the $(N+1)$th eigenstate.   

We further define a factor $Q$ based on the butterfly-shaped spectral structure, which reveals a linear relationship between the spectral distribution of 1D PQCs and irrational numbers. 
This relationship depends solely on the size of $\beta$ (not near $0$ or $1$) and is independent of its type. 
Specifically, when $\beta < \beta_c$, $Q = 1 - \beta$; and when $\beta > \beta_c$, $Q = \beta$. The linear relationship disappears as the irrational number $\beta$ approaches $0$ or $1$. 

These results further our understanding of the spectral structure of 1D PQCs and highlight the crucial role of irrational numbers. 
We also realize that our understanding of the spectral structure of 1D PQCs systems is still not comprehensive enough. 
For example, how does the number of eigenvalues change as the computational resolution $N$ increases? What is the spectral distribution when two irrational numbers exist in a QOL? These questions await further exploration in the future.

\begin{acknowledgments}

The authors acknowledge financial support from the National Key R\&D Program of China (2023YFA1008802), the Science and Technology Innovation Program of Hunan Province (2024RC1052), the Innovative Research Group Project of Natural Science Foundation of Hunan Province of China (2024JJ1008), the Postgraduate Scientific Research Innovation Project of Hunan Province (CX20240592), and the National Natural Science Foundation of China (12501603). We are also grateful to the High Performance Computing Platform of Xiangtan University for partial support of this work.

\end{acknowledgments}



\bibliography{references_1}

\end{document}